# Early monsoon drought and mid-summer vapor pressure deficit induce growth cessation of lower margin *Picea crassifolia*

**Running head** Growth cessation corresponds to drought.

SHOUDONG ZHAO[1], YUAN JIAN[1*], MANYU DONG[1], HUI XU[2], NEIL PEDERSON[3]

[1]Beijing Key Laboratory of Traditional Chinese Medicine Protection and Utilization, Faculty of Geographical Science, Beijing Normal University, 19 Xinjiekouwai Street, Haidian, 100875 Beijing, China

[2]Department of Biostatistics and Epidemiology, School of Public Health and Health Sciences, University of Massachusetts Amherst, 131 Brittany Manor Dr. APT G, Amherst, MA 01003, USA

[3]Harvard Forest, Harvard University, 324 N Main Street, Petersham, MA 01366, USA

---

* Corresponding author. E-Mail: jiangy@bnu.edu.cn. Tel: +86 10 58806093.

**Abstract**

Extreme climatic events have been shown to be strong drivers of tree growth, forest dynamics, and range contraction. Here we study the climatic drivers of *Picea crassifolia* Kom., an endemic to northwest China where climate has significantly warmed. *Picea crassifolia* was sampled from its lower distributional margin to its upper distributional margin on the Helan Mountains to test the hypothesis that 1) growth at the upper limit is limited by cool temperatures and 2) is limited by drought at its lower limit. We found that trees at the lower distributional margin have experienced a higher rate of stem-growth cessation events since 2001 compared to trees at other elevations. While all populations have a similar climatic sensitivity, stem-growth cessation events in trees at lower distributional margin appear to be driven by low precipitation in June as the monsoon begins to deliver moisture to the region. Evidence indicates that mid-summer (July) vapor pressure deficit (VPD) exacerbates the frequency of these events. These data and our analysis makes it evident that an increase in severity and frequency of drought early in the monsoon season could increase the frequency and severity of stem-growth cessation in *Picea crassifolia* trees at lower elevations. Increases in VPD and warming would likely exacerbate the growth stress of this species on Helan Mountain. Hypothetically, if the combinations of low moisture and increased VPD stress becomes more common, the mortality rate of lower distributional margin trees could increase, especially of those that are already experiencing events of temporary growth cessation.

**Introduction**

In recent years, the intensity and frequency of extreme climatic events, such as heat waves and drought, have changed (Sun et al., 2014; Wang and Zhou, 2005; Zhang and Zhou, 2015). Extreme climatic events have been shown to be strong divers of forest growth, dynamics, and contraction. For example, extreme heat and drought in 2013 led to about 40% reduction in gross primary production in Southern China (Yuan et al., 2016). Although trees can recover from growth declines (Anderegg et al., 2015; Gazol and Camarero, 2016; Orsenigo et al., 2014; Young et al., 2017), some do not, resulting in forest decline and rising tree mortality rates (Allen et al., 2015; Zhang et al., 2014). Tree mortality events change forest structure and composition and result in species' range and landscape shifts in severe conditions (Allen and Breshears, 1998; Gray et al., 2006; Martínez-Vilalta and Lloret, 2016). Understanding tree growth response to extreme events is important for forest conservation and management in response to ongoing climate change.

The potential impact of synergistic extreme events, such as hotter droughts, on trees is uncertain under future climate change (Allen et al., 2015). Research indicates that warming has sped up ecological processes in some regions. Forest inventory, remote sensing, dendrochronological data, and a combination of forest census approaches have all indicated that warming has induced some shifts at northern range margins (Chen et al., 2011; Mathisen et al., 2014; Shrestha et al., 2015). In contrast, prior research finds that tree growth within the southern distribution or at southern range margins of some species is often limited by a lack of soil water and exacerbated by high temperatures (Galván et al., 2014; Hacket-Pain et al., 2016; Martin-Benito and Pederson, 2015; Tiwari et al., 2017; Urli et al., 2015). A few studies have found evidence of a decline in growth and increased mortality with warming (Kharuk et al., 2013; Lesica and McCune, 2004; Peñuelas et al., 2008). From these findings, we can infer that interactive effects of warming and extreme events on trees are region-specific, especially at different range margins.

*Picea crassifolia* Kom. (Qinghai spruce) is an endemic tree species in semi-arid northwest China that could be significantly impacted by extreme climatic events. *Picea crassifolia* growth is sensitive to drought and appear to broadly have a similar climate-growth relationship across elevation (Fang et al., 2009; Gao et al., 2013; Gou et al., 2005; Liang et al., 2016; Zhang et al., 2017). An important harbor for *Picea crassifolia* are the isolated forests of China's Helan Mountains. A little more than 5% of the total *Picea crassifolia* biomass can be found in these mountains, a total that is second only to Qilian Mountains (Liu, 1992). Embedded within the continental transition from steppe to desert vegetation (Jiang et al., 2007), the Helan Mountains represent the northeastern margin of the *Picea crassifolia* distribution area. As a result, the Helan Mountains are an important region for *Picea crassifolia* conservation in the face of climatic change.

Here, we investigate the potential climatic drivers of *Picea crassifolia* growth across its elevational distribution in the Helan Mountains. The only investigation of climate-growth patterns of *Picea crassifolia* in the Helan Mountains found that dry summers and cool October temperatures prior to the growing season are the primary climatic constraints on growth (Zhang et al., 2009). Zhang et al. (2009) study represents the climate-growth relation of *Picea crassifolia* at one specific band of forest in the Helan Mountains, not across its elevational distribution. Because of this, we sampled *Picea crassifolia* at its upper, middle, and lower elevational distributional margins along an eastern slope. As elevation increases in the mountains of China, temperature decreases and precipitations increases (Fu, 1995). Thus, our first hypothesis is that *Picea crassifolia* growth at its upper distributional margin significantly and positively responses to temperature. Our second hypothesis is that *Picea crassifolia*'s growth at its lower margin in the Helan Mountains significantly and positively responses to moisture availability. For the second hypothesis, statistically significant and positive correlations to precipitation and a statistically significant, but negative correlations to maximum temperatures and vapor pressure deficit (VPD) would be interpreted as a positive relationship between *Picea crassifolia*'s growth and moisture availability.

**Materials and methods**

*Study site*

Our study region is located on the middle part of the Helan Mountain, Ningxia Hui Autonomous Region, China (Fig. 1). The Helan Mountains region is also the northwestern extent of the East Asian summer monsoon (Wang and Lin, 2002) and located at arid and semi-arid areas in Inner Asia. There are two belts of coniferous forests across elevation within this region: one is dominated by *Pinus tabulaeformis* Carr. from 1900-2350 m elevation while the other one is dominated by *Picea crassifolia* above 2350 m (Jiang et al., 2007). Meteorological data from the Alpine meteorological station from 1961-1990 indicates that the mean annual temperature is -0.7 °C while the mean monthly growing season temperature (May-September) is above 5 °C (Fig. 2). Annual total precipitation is 427 mm, 40% of which occurs during the peak of the monsoon season, from July to August. Overall, annual temperatures and vapor pressure deficit have been increasing while total annual precipitation has been relatively stable over time (Fig. 3). Variability in early growing season precipitation, e.g. May and June, has increased in recent decades.

*Tree-ring data*

Increment cores of *Picea crassifolia* were collected from three sites along an elevation gradient, named the lower distributional margin (LDM), mid-distribution (MID), and what we found to be an upper distributional margin (UDM) on this transect (Table 1). All sites were located on slopes of 15-25 degree. At each site, 25 mature and healthy-appearing canopy trees were selected for sampling. At least two cores were extracted from each tree at breast height (1.3m from the ground), though some trees had three or four cores extracted as coring occasionally went through the entire stem. Samples were dried, mounted, and sanded with progressive finer sand paper. Samples were visually cross-dated to identify locally-absent rings and false rings where necessary (Stokes and Smiley, 1968). Ring widths were then measured to the nearest 0.01 mm using a LINTAB 5 measuring table (RINNTECH, Heidelberg, Germany) with the accuracy of cross-dating was checked by program COFECHA (Holmes, 1983). Cores that were too rotten to accurately date were not used for this study.

Because we analyzed the climate-growth relationship at the tree level, standardization was conducted in a way that preserved as much low frequency information as possible. If there was no evidence of growth releases related to change in competition (Lorimer, 1985) in individual trees, for example, size-related trends in tree growth were removed through the use of negative exponential functions. Trees with trends that indicated release from competition or other ecological processes were detrended using 30-year smoothing splines. Standardized ring-width series from the same tree were averaged into single-tree chronologies. Similarly, all individual single-tree chronologies from the same site were averaged into site-level chronologies using bi-weight robust mean (Cook, 1985).

Basal area increment (BAI) is used to detect long-term growth trend (Peters et al., 2015). Basal area increment for every decade in each series was calculated as

$$BAI_t = \pi(R_t^2 - R_{t-1}^2) \qquad (1)$$

where $R$ is the stem radius and $t$ is the decade of tree rings formation. BAI from the same tree were averaged into single-tree BAI by decade. We recognize that by going back in time with BAI, there are significant uncertainties on the trends in BAI due to the loss of trees adjacent to the surviving trees that were sampled going back in time, i.e. ghost trees (Babst et al., 2014; Foster et al., 2014; Nehrbass-Ahles et al., 2014). New research, however, indicates that estimating biomass in forests that are essentially even-aged and in the stem-exclusion stage might have enough certainty to be useful going back almost five decades (Dye et al., 2016).

*Climate data*

Meteorological data comes from three stations near the study area (Fig. 1): Alpine (AL), Alxa Left Banner (ALB), and Yinchuan (YC). AL was the nearest station to the sample sites, but provided only 30-year data (1961-1990). ALB and YC both provided meteorological data from 1950s to present. Thus, the average annual values for AL were used to describe the climate of the study area while ALB and YC were used to describe the climate variation. Monthly data of precipitation, minimum temperature, maximum temperature, wind speed, hours of sunshine, relative humidity, and surface atmospheric pressure was used to estimate 1-month SPEI (SPEI-1) via the FAO56 Penman-Monteith method (Allen et al., 1998). Daily data of minimum temperature, maximum temperature, and relative humidity was used to estimate vapor pressure deficit (VPD, Allen et al., 1998). Daily VPD values were averaged to calculate monthly mean VPD. While the site location of our meteorological stations could introduce some uncertainty in our findings, monthly precipitation, minimum temperature, maximum temperature, SPEI-1, and VPD from ALB and YC were averaged to avoid single-station bias and reflect regional climate change conditions (Blasing et al., 1981).

*Statistical analysis*

Two statistics were used to estimate the potential climatic effect on tree growth, the standard deviation (SD) and the first-order autocorrelation coefficient. A higher SD suggests a stronger climatic influence on tree growth while a higher first-order autocorrelation coefficient suggests a stronger influence of tree growth from one year to the next (Cook, 1985). Both SD and first-order autocorrelation coefficient were calculated for the common intervals of all single-tree chronologies.

The limiting factors of radial growth were identified by correlation analysis. Pearson's correlation coefficients were calculated between single-tree chronologies and climate variables from 1954-2010. Climate factors included total monthly precipitation, monthly minimum, and monthly maximum temperature, SPEI-1, and monthly VPD from previous May to current October to cover the growing season (May-September) of *Picea crassifolia* (Liang et al., 2006) and include lags due to non-structural carbon.

First-order autocorrelation coefficients and correlation coefficients was converted into a Z-score using Fisher's *r*-to-*z* transformation for comparing. One-way ANOVA followed by Tukey HSD test was used to test difference in tree age, diameter at breast height (DBH), SD, first-order autocorrelation coefficients, and correlation coefficients between our three sites.

**Results**

*Occurrence of locally-absent rings*

The frequency of locally-absent rings is greater at the lower elevation (LDM) than at higher elevations (Table 2). Locally-absent rings occur in more years at LDM than at MID and UDM (12 versus 6 and 4 years, respectively). Even when the trees were younger, during 1940s for example, locally-absent rings were only documented at the LDM site. Of the trees with locally-absent rings in three or four radii, 95% are growing at the LDM. Nearly all LDM trees, 92%, have locally-absent ring in at least one radius over the lifetime of the tree, while 68% have locally-absent rings in at least two radii and 44% in at least three radii.

In recent decades, high rates of locally-absent rings occurred in four years, 1982, 2001, 2005, and 2008. Again, the LMD site had the highest rate of locally-absent rings during these years. The percentage of LMD trees with locally-absent rings during these years ranged from 48-84%, while the ranges for MID and UDM was 0-12% at each site. During these four years at LDM, 12-36% of trees had locally-absent rings in at least three radii.

The rates of locally-absent rings seemed high for spruce trees (genus *Picea*). To understand if these rates are unusually high, we conducted an analysis of the rate of locally-absent rings of spruce archived in the International Tree-Ring Databank (https://www.ncdc.noaa.gov/data-access/paleoclimatology-data/datasets/tree-ring); we could not find rates of locally-absent rings of spruce in the St. George et al. (2013) review. Labeling of ITRDB samples made it hard to calculate the rate of locally-absent rings at the tree level such that we calculated maximum rates at the radii level per site for this analysis. Of the 797 spruce sites in the ITRDB version 7.13, 605 have 20 radius or more (assuming to contain roughly 10-20 trees). Of these collections in the ITRDB that appear to have an equal number of trees and radii sampled in our sites, only one collection (Switzerland, latitude: 46.30 °N, elevation: 840 m) has more locally-absent rings than our LDM site (Fig. 4),.

*Tree growth and age and size structure analysis*

Mean tree age at LDM and UDM are slightly higher than that at MID (89 and 93 years versus 84 years, respectively), but there is no significant difference in tree age between each location ($F = 1.769$, $p = 0.178$; Fig. 5). Similarly, while mean DBH differ across sites (LDM: 21 cm, MID: 22 cm, UDM: 23 cm), there is no significant difference in mean DBH ($F = 1.684$, $p = 0.193$; Fig. 5).

The standard chronologies have a similar amount of high frequency variation in radial growth across elevations (Fig. 6); they crossdate. There is, however, a significant difference of SD between the elevations sampled ($F = 115.8$, $p < 0.001$; Fig. 5). The mean SD at LDM is significantly higher than at MID and UDM (0.605 versus 0.415 and 0.390, respectively), indicating LDM trees could be more sensitive to climatic variation compared to the other locations. There is also a significant difference of first-order autocorrelation coefficients between the three elevations ($F = 56.19$, $p < 0.001$; Fig. 5). The mean first-order autocorrelation coefficient at LDM is nearly zero and significantly less than at MID and UDM (-0.025 versus 0.303 and 0.247 respectively). These values indicate that trees at LDM have a lower persistence structure (autocorrelation from one year to the next) and that climate variation potentially has a greater influence on annual tree growth at lower elevation. The variation of the LDM chronology has increased in recent decades (SD of the site-level standard chronology before 1990 is 0.276, and that after 1991 is 0.463). At the same time, the mean decadal biomass accumulation of *Picea crassifolia* trees at their LDM has significant decreased from the 1970s (43.3 cm$^2$) to the 2000s (20.9 cm$^2$; Fig. 6).

*Climate-growth relationships*

The correlations between climate and single-tree chronologies are similar between sites (Fig. 7). Generally, growth is positively correlated with total May and total June precipitation, and previous September, March, May, and June SPEI. Growth is also negatively correlated with average March and average July temperature, and average previous September, average March, average May, and average July VPD.

Single-tree chronologies at LDM have significant higher correlation coefficients than those at MID and UDM for total May and total June precipitation. High rates of locally-absent rings at LDM (50% and greater) are most strongly correlated with low June precipitation (Fig. 8). Years with rates of locally-absent rings higher than 25 cluster more tightly when both low June precipitation and high July VPD occur; that is, the combination of low June precipitation and high July VPD helps to better account for climatic drivers of high rates of locally-absent rings (lower left panel in Fig. 8).

**Discussion**

*Cessation of stem growth in Picea crassifolia*

Our research indicates that, despite all populations having a similar structure in monthly climatic sensitivities, 1) the rate of locally-absent rings of *Picea crassifolia* was greater at its lower distributional margin site versus populations at higher elevations on the eastern slope of Helan Mountain; 2) the rate of severe, locally-absent ring events in canopy trees has increased since 2001; and 3) the lower distributional margin population had the second highest rate of locally-absent rings when compared to the current collections of *Picea* with a comparable sample replication in the Northern Hemisphere.

'Locally-absent ring' is dendrochronological term indicating that a tree had no visible or measurable growth in the sampled radius during a specific year. Ecologically speaking, a locally-absent ring represents a cessation of stem growth along the radius examined. These events can only be identified through the crossdating of radial growth samples (from cores or cross-sections), a process that identifies the pattern of common large and small rings within a tree and populations (Black et al., 2016; Douglass, 1920; Fritts, 1976; Stokes and Smiley, 1968). Little to no stem growth during a specific year, or, potentially, a complete lack of radial growth around the stem would be difficult to detect quantify with great certainty if the basic principles of dendrochronology are not followed. We find that rates of locally-absent rings at low elevation *Picea crassifolia* are the second highest for *Picea* of the 605 large populations that have been sampled to date around the Northern Hemisphere.

The temporary cessation of stem growth is one way trees adjust to extreme conditions. A few common factors of locally-absent rings include extreme climate (Douglass, 1920), insect outbreaks (Alfaro et al., 1982), severe fire (Zackrisson, 1980), the competitive stress placed on understory trees by canopy trees (Lorimer et al., 1999), and earthquakes (Jacoby, 1997; Jacoby et al., 1988). Fire is rare in our study region, insects are not typically an ecological factor of this species, and the *Picea crassifolia* trees we sampled are not on an earthquake fault zone (Geng and Shan, 1992). Because we sampled overstory trees, it is difficult to see how canopy position accounts for the differences in locally-absent ring rates between the lower distributional margin trees versus trees at higher elevations. All trees sampled here were in the canopy position. Also, we found no significant difference in the ages or diameters between site, indicating each sites was in a similar stage of forest development. We do not have measures of stand density, but because tree sizes and ages are so similar, it is reasonable to think stand densities are not greatly different. We cannot, however, rule out stand density as a factor in the differing rates between sites. Given this, it seems most likely that low precipitation early during the monsoon season and mid-summer dryness are the drivers of the increase in stem-growth cessation events (Figs. 7, 8).

*Drought effects on tree growth*

We found that *Picea crassifolia* growth is sensitive to moisture availability across an elevation gradient in the Helan Mountains. Specifically, moisture availability here is expressed as a signficant response to low precipitation and high temperatures and VPD. These results do not support our first hypothesis that the population at the highest elevation would be limited by cool temperatures. Climatic-sensitivity analyses do, however, support our second hypothesis, that lower distribution populations are limited by reduced moisture availability.

The effect of low precipitation early during the monsoon season on the populations of *Picea crassifolia* we sampled on Helan Mountains support the relation between climate and *Picea crassifolia* growth over much of its distribution (e.g. Fang et al., 2009; Gao et al., 2013; Gou et al., 2005; Liang et al., 2016; Zhang et al., 2017). Only at a few high elevation sites does warming enhance *Picea crassifolia* growth (Chen et al., 2012). We are not yet sure if these relations between low precipitation and high temperatures in Helan Mountains *Picea crassifolia* are consistent through time. A study investigating a changing climatic sensitivity is an important next step in this line of research.

An important aspect of future research would be to examine the subtle, but seemingly a hypothetical nonlinear impact of moisture conditions on *Picea crassifolia* growth over elevation on Helan Mountains. Even though each population had the same basic climatic response, strength of chronology statistics (SD, first-order autocorrelation coefficients), and calculated climatic sensitivity, the differences, while smaller, seemed to have a larger impact moving from the lower elevation to the mid and upper elevations: the rate of locally-absent rings at the lower margin are significantly greater when compared to populations at higher elevations. The impact of moisture availability on growth is different moving from the lower boundary to interior of *Picea crassifolia* forest follows a general phenomenon in other studies (Cernusak and English, 2015). This is still, however, an active line of research that needs further research. European beech, for example, appears to be more vulnerable to climate in the middle of its range in central Europe versus range margins (Cavin and Jump, 2017; Hacket-Pain et al., 2016).

*Implications of Picea crassifolia growth and climatic sensitivity*

Hypothetically, the increase in temporary cessation of stem growth events found in our study is an important ecological signal. While it is not possible to extrapolate these events detected at the radii scale across the stem, a high frequency of reduced growth might be a signal of increased tree stress. In fact, an increase in the frequency of locally-absent rings is sometimes found prior to a decline in radial growth or tree mortality (Cailleret et al., 2017; Liang et al., 2016). Because the lower margin trees in our study had the highest rate of stem-growth cessation events, that the rates in this population have increased over the last two decades, and that three of the four most severe events have occurred since 2001, it stands to reason that this population has become more stressed over the last two decades. Further supporting this conclusion is that biomass (BAI) declined significantly in 2000s compared to prior decades.

Based on 1961-1990 averages from the AL meteorological station, May through July total potential evapotranspiration often approaches that of total precipitation (162.2 mm of total potential evapotranspiration via the Thornthwaite method versus 165.6 mm of total precipitation). Thus, when reduced rainfall or increased maximum temperatures occur, it is likely that soil water is not sufficient for photosynthesis. In fact, during the four years with the most intense stem-growth cessation events (1982, 2001, 2005, and 2008), total June precipitation was 0.7-1.3 SD below the mean (Fig. 8). In 2005, when 84% of the sampled trees had at least one locally-absent ring, total June precipitation was 1.3 SD below the mean. The evidence here indicates that low precipitation in June, just prior to the peak in monsoonal precipitation, is likely the most common trigger of increasing locally-absent ring events since 2001.

To date, there has been little to no trend in the amount of precipitation within the distribution of *Picea crassifolia*. Changing temperatures, however, could have increased the severity of drought in recent decades–warming is expected to increase dry conditions globally over the next several decades (Cook et al., 2014). Since 1960, temperature has significantly increased across the distribution of *Picea crassifolia* (Piao et al., 2010), resulting in increased potential evapotranspiration, reduced soil moisture, and an increase in soil drought events (Wang et al., 2011). These changes can impact tree growth in two ways. First, growth declines or ceases during severe drought. Second, plants in arid areas need several years to recover from drought events (Anderegg et al., 2015). Thus, an increase in precipitation variation accompanied by warming could be increasing the moisture stress on *Picea crassifolia*. In fact, we have found that higher vapor pressure deficits in July are likely an important additional factor in the cessation of stem growth of *Picea crassifolia* in the Helan Mountains (Fig. 8). Similar interaction of trends in moisture, temperature and other environmental factors will cause complex vegetation responses of Inner Asia (Liu et al., 2013; Poulter et al., 2013).

We hypothesize that increased warming without a substantial increase in precipitation would induce a significant decline in growth and, possibly, an increase in mortality of lower distribution *Picea crassifolia* in the Helan Mountains. Similar findings have been observed in other semi-arid forests (Settele et al., 2015 and references within). It is hard to predict how the *Picea crassifolia* forest on the Helan Mountains might change given the uncertainty in forecasting future precipitation. It might be that other species increase in importance during warming as these species become a better overall competitor versus *Picea crassifolia* at lower elevations. If so, the community structure could change, resulting in a narrowed elevational distribution of *Picea crassifolia*. Alternatively, cooling and wetting would presume to have a near opposite effect. These hypothetical scenarios around the Helan Mountain *Picea crassifolia* populations are ripe for modeling experiments. Within the entire distribution of *Picea crassifolia*, however, there is substantial uncertainty around this potential trajectory of forest development. *Picea crassifolia* in the Qilian Mountains, for example, has more stem-cessation events at upper and middle elevations than lower elevations. Differences in canopy cover and stem density appear to be at work in the Qilian Mountains populations (Liang et al., 2016). That is, it seems that high intraspecific competition in *Picea crassifolia* in the Qilian Mountains populations enhances climate effects on tree mortality and growth decline. At the very least, our results in this study fall in line with observations that extreme heat and drought events reduce primary productivity (Liu et al., 2013; Poulter et al., 2013), suggesting that an increase in extremely hot years or severe drought could reduce the ability of the *Picea crassifolia* forest on the Helan Mountains to act as a carbon sink.

**Conclusions**

We found that the growth of *Picea crassifolia* across elevation in the Helan Mountains of northwest China is driven by moisture availability. Trees at a lower distributional margin, however, appear to have reacted more strongly to reduced precipitation than those at higher elevations through the significant increase in stem-growth cessation in recent decades. A significant reduction in early-monsoonal precipitation appears to the primary cause of cessation in stem growth in lower distributional margin trees with increased mid-summer vapor pressure deficit exacerbating low precipitation conditions. In other types of ecosystems, it has been shown that stem-growth cessation often precedes increased rates of tree mortality. If future environmental conditions deteriorate, climatic change could become a threat to the endemic *Picea crassifolia* on the Helan Mountains.

**Acknowledgements**

This work was supported by the National Natural Science Foundation of China (NSFC No. 41630750). We are very grateful to Jingfan Wang and Hui Wang for their help during fieldwork.

**Table 1.** Characteristics of the sampling sites.

| Site | Latitude (° N) | Longitude (° E) | Elevation (m a.s.l.) | Main aspect | Slope (°) | No. trees/radii |
|------|----------------|-----------------|----------------------|-------------|-----------|-----------------|
| UDM  | 38.77          | 105.90          | 2886                 | Northeast   | 25        | 25/78           |
| MID  | 38.77          | 105.90          | 2670                 | North       | 25        | 25/89           |
| LDM  | 38.78          | 105.90          | 2408                 | Northeast   | 15        | 25/90           |

**Table 2** Years with locally-absent rings since 1941.

| Year | Rate of trees with locally-absent rings in at least one radius | | |
|------|-----|-----|-----|
|      | LDM | MID | UDM |
| 1941 | 4   | 0   | 0   |
| 1947 | 8   | 0   | 0   |
| 1973 | 0   | 0   | 4   |
| 1981 | 8   | 0   | 4   |
| 1982 | 52  | 12  | 12  |
| 1991 | 12  | 0   | 4   |
| 1994 | 20  | 0   | 4   |
| 1995 | 4   | 0   | 0   |
| 2000 | 8   | 0   | 0   |
| 2001 | 60  | 4   | 0   |
| 2005 | 84  | 4   | 4   |
| 2006 | 24  | 4   | 0   |
| 2008 | 48  | 0   | 0   |

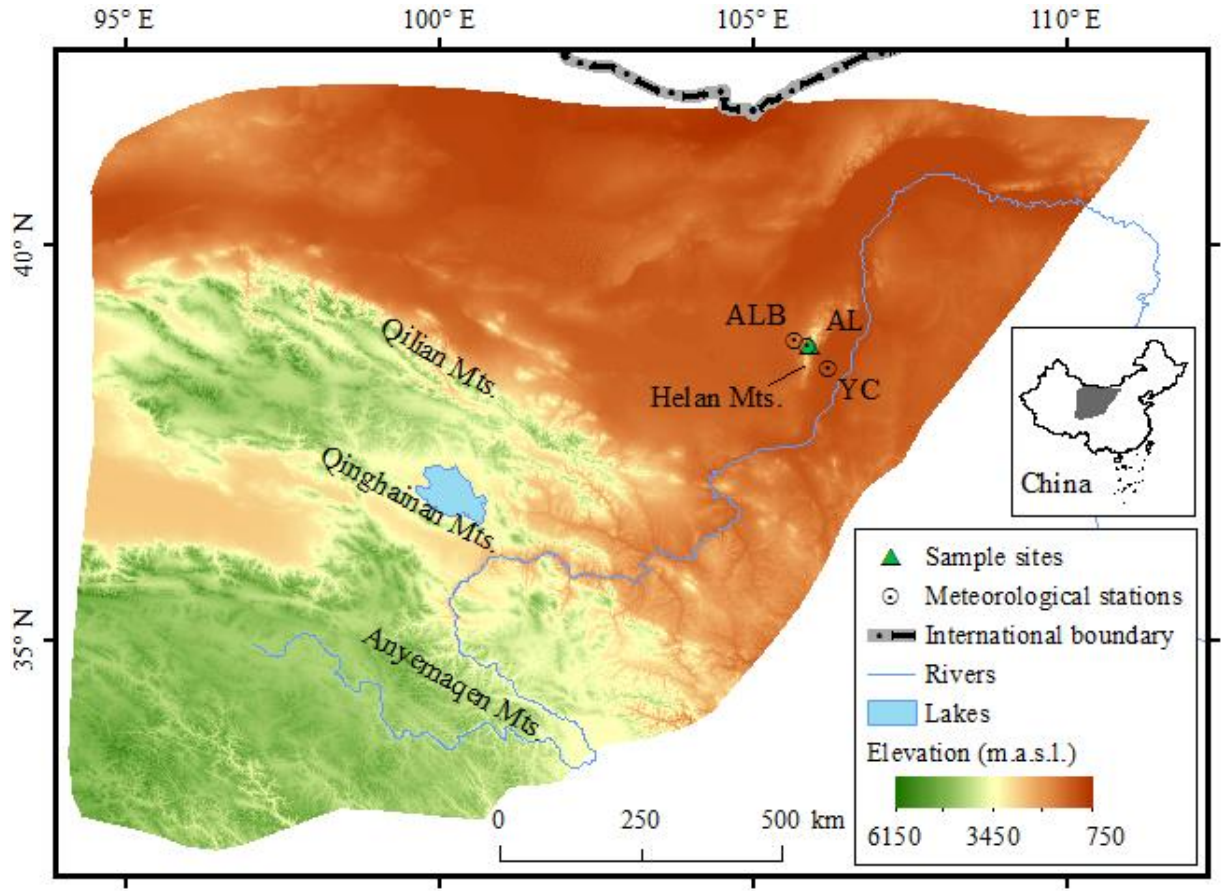

**Figure 1.** Sample sites, meteorological stations, and *Picea crassifolia* spatial distribution (modified from Liu, 1992).

**Figure 2.** Climate diagrams (after Walter, 1979) from Alpine (AL), Alxa Left Banner (ALB), and Yinchuan (YC) stations (locations are shown in Figure 1). Dashed lines show the mean monthly temperature, solid lines the mean monthly precipitation sum. Gray areas correspond to relative dry periods.

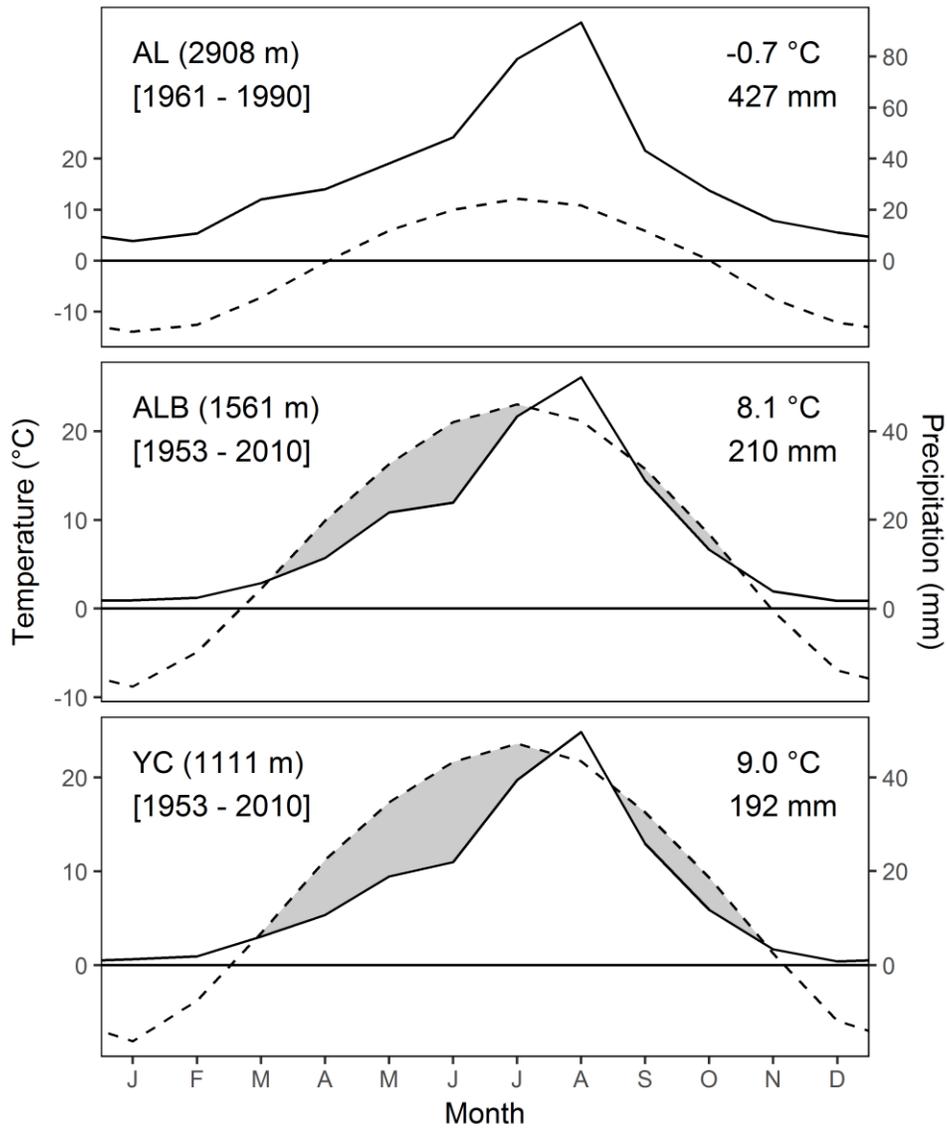

**Figure 3.** Time series (solid lines) and linear trends (dash lines) of annual meteorological parameters.

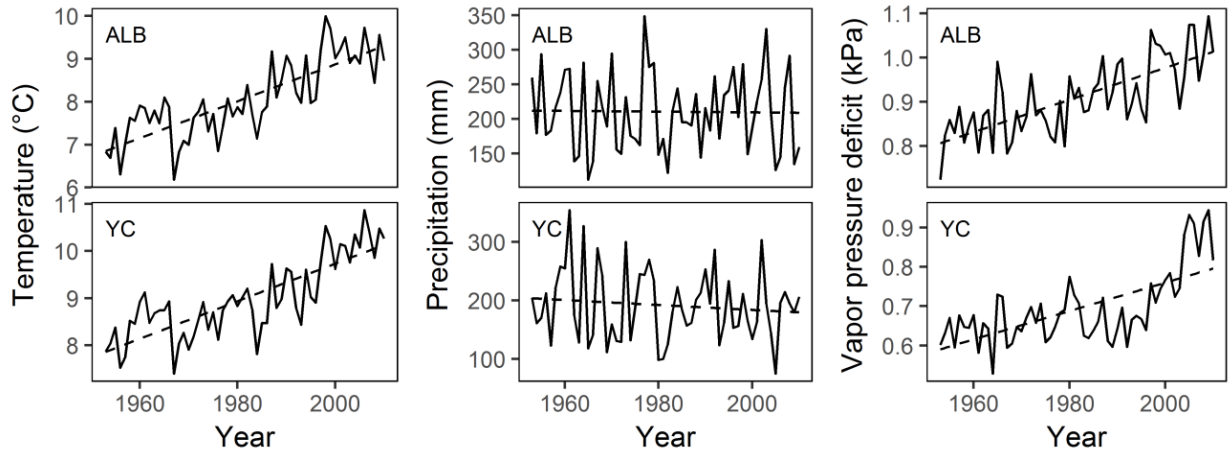

**Figure 4.** Maximum rate of locally-absent rings of spruce collections from the ITRDB with 20 series or more (circles) versus our collections in the Helan Mountains (squares). Y-axis is in log scale. Only collections whose maximum rates of locally-absent rings are larger than 0 are shown.

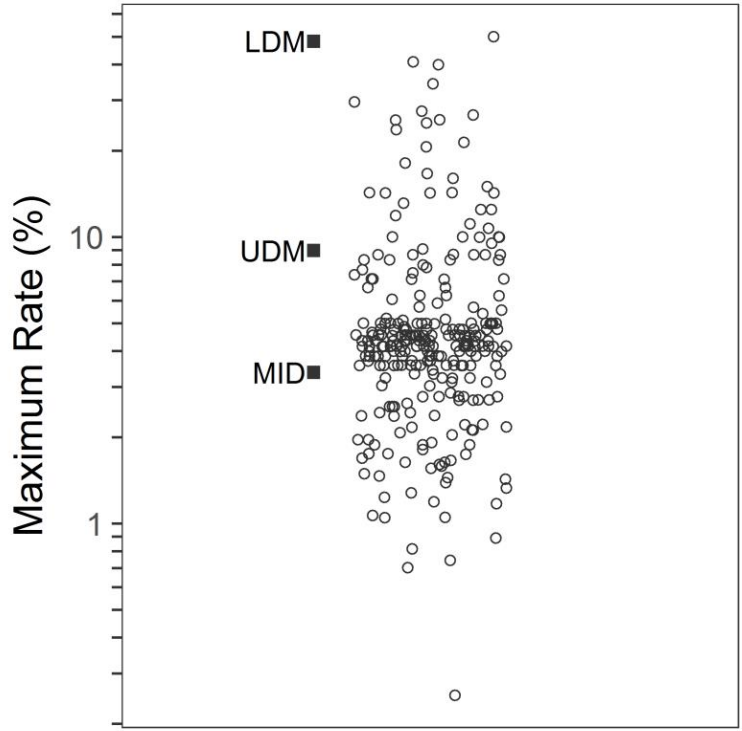

**Figure 5.** Tree age, diameter at breast height (in cm), and characteristics of single-tree standard chronologies. Boxes show the median, 25th and 75th percentiles, whiskers extend to the minimum and maximum, and letters indicate differences between sites (Tukey HSD test after ANOVA).

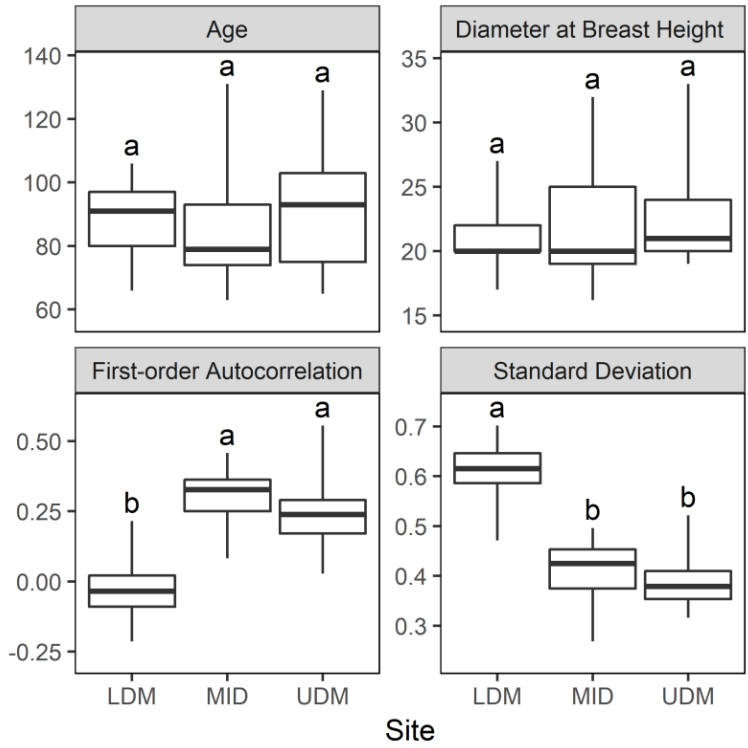

**Figure 6.** Single-tree and site-level standard chronologies (gray and black lines respectively), and single-tree basal area increment by decade (boxplots). Boxes show the median, 25th and 75th percentiles, whiskers extend to the minimum and maximum, and letters indicate differences between decades (Tukey HSD test after ANOVA).

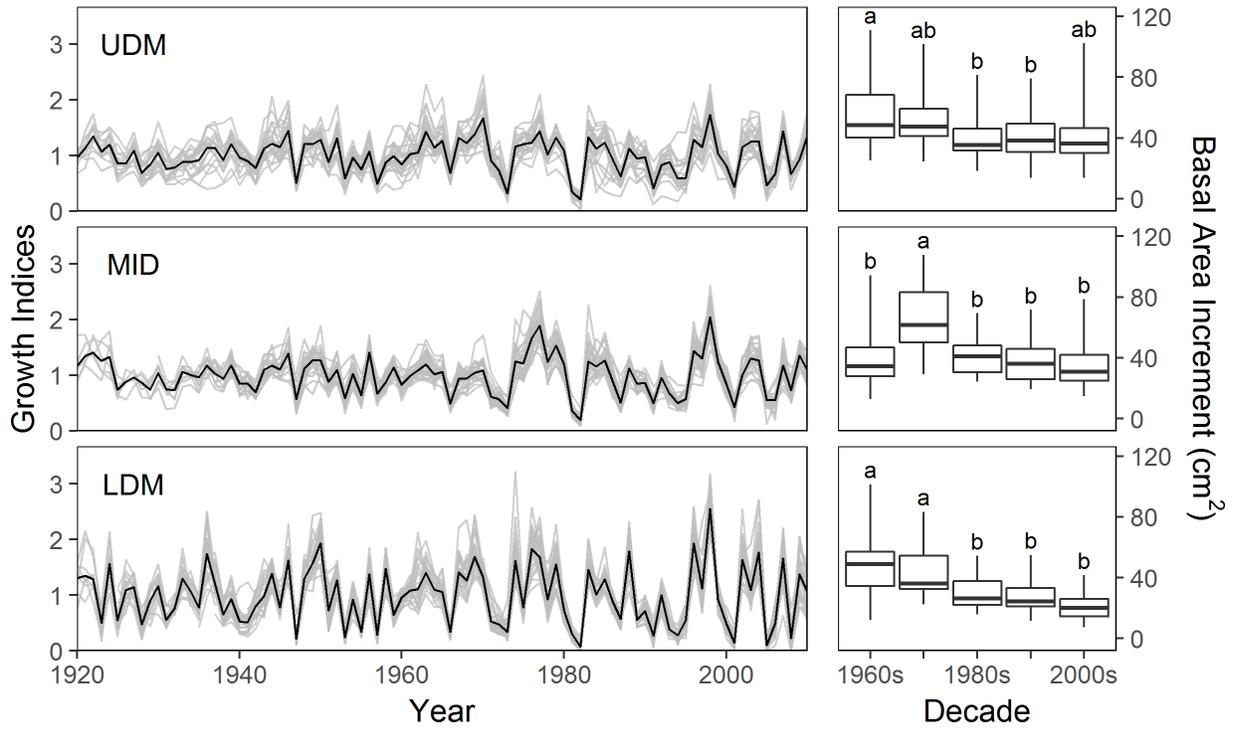

**Figure 7.** Correlations coefficients between climate variables and single-tree chronologies from LDM (red), MID (green), and UDM (blue). Boxes show the median, 25th and 75th percentiles, and whiskers extend to the minimum and maximum. Lowercase letters represent months of the previous year, and uppercase letters represent months of the current year.

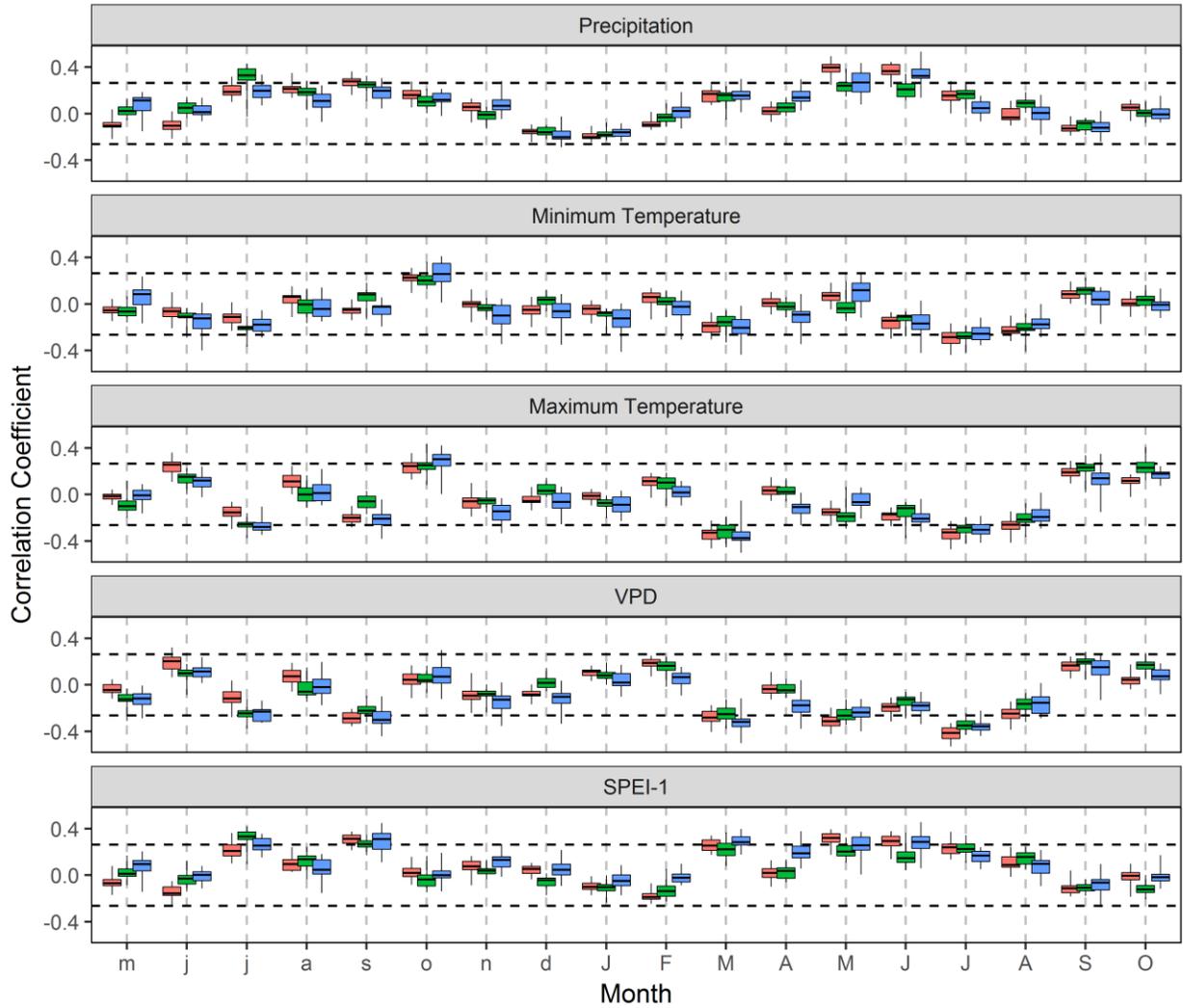

**Figure 8.** Relationship between rate of trees with locally-absent rings in at least one radius at LDM and climatic variables most strongly related to radial growth. Note how the years with high rates of locally-absent rings cluster as July vapor pressure deficit (VPD) is added to June precipitation (lower right).

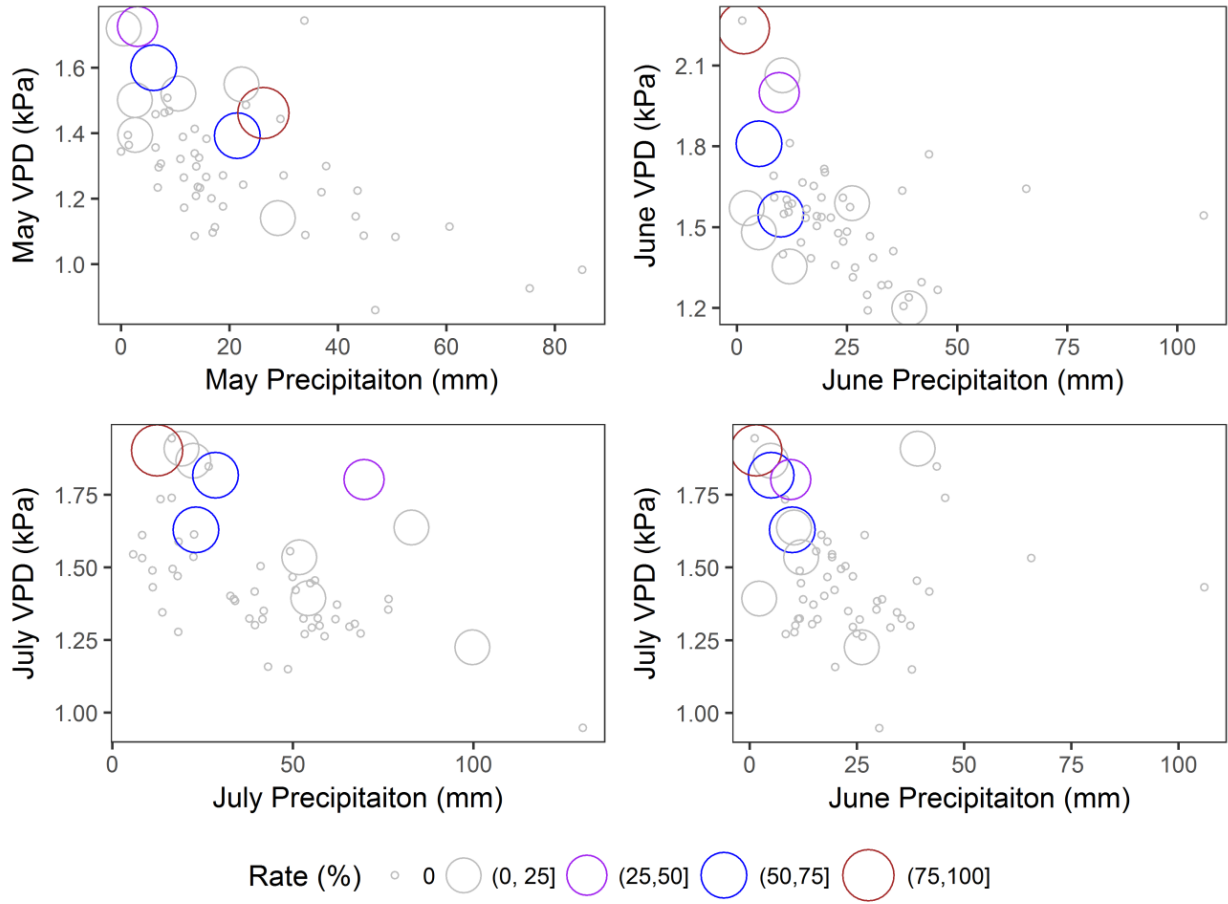